\documentclass[sigconf, nonacm, pdfa]{acmart}

\usepackage{xspace}

\newcommand\vldbdoi{XX.XX/XXX.XX}
\newcommand\vldbpages{XXX-XXX}
\newcommand\vldbvolume{14}
\newcommand\vldbissue{1}
\newcommand\vldbyear{2020}
\newcommand\vldbauthors{\authors}
\newcommand\vldbtitle{\shorttitle} 
\newcommand\vldbavailabilityurl{https://github.com/curtis-sun/LLM4Rewrite}
\newcommand\vldbpagestyle{plain}

\usepackage{graphicx}
\usepackage{balance}  % for  \balance command ON LAST PAGE  (only there!)
\usepackage{epsfig}
\usepackage{epstopdf}
\usepackage{latexsym}
\usepackage{enumitem}
\PassOptionsToPackage{noend}{algpseudocode}
\usepackage{algpseudocode}
\usepackage{pifont}
\usepackage{epsfig}
\usepackage{amssymb}
\usepackage{amsmath}
\usepackage{amsfonts}
\usepackage{stmaryrd}
\usepackage{url}
\usepackage{multirow}
\usepackage{array}
\usepackage[normalem]{ulem}
\usepackage{color}
\usepackage{cleveref}
\usepackage{xspace}
\usepackage{mathtools}
\usepackage{soul}
\usepackage{listings}
\usepackage{xcolor}
\usepackage{tikz}
\newsavebox{\blackball}
\newsavebox{\greenball}

\usepackage[utf8]{inputenc}

\usepackage{bbding}
\usepackage{endnotes}
\usepackage{booktabs}
\usepackage{threeparttable}
\usepackage{tabulary}
\usepackage{CJKutf8}
\usepackage[most,breakable]{tcolorbox}
\usepackage{etoolbox}
\usepackage{fancyhdr}
\usepackage{lipsum}
\usepackage[fencedCode]{markdown}
\usepackage{marginnote}
\usepackage{makecell}

\usepackage{subcaption}

\usepackage[lined,boxed,vlined,ruled,linesnumbered]{algorithm2e}

\newcommand{\hi}[1]{\vspace{.25em} \noindent {\bf #1}\xspace}

\newcommand{\daagent}{\textit{DA-Agent}\xspace}
\newcommand{\claudecode}{\textit{Claude Code}\xspace}
\newcommand{\oursys}{\texttt{ACID-Agent}\xspace}

\usepackage{tabularx} % 加载tabularx宏包

\begin{document}

\title{Agentic Transaction: Towards ACID-Compliant Agent Systems}

%%
%% The "author" command and its associated commands are used to define the authors and their affiliations.
\author{Zhaoyan Sun}
\affiliation{%
  \institution{Tsinghua University}
  % \city{Beijing}
  % \country{China}
}
\email{szy22@mails.tsinghua.edu.cn}

\author{Xiaoxiao Wang}
\affiliation{%
  \institution{Tsinghua University}
  % \city{Ithaca}
  % \country{United States of America}
}
\email{xw724@cornell.edu}

\author{Guoliang Li}
\affiliation{%
  \institution{Tsinghua University}
  % \city{Beijing}
  % \country{China}
}
\email{liguoliang@tsinghua.edu.cn}

\pagestyle{plain}

\pagenumbering{arabic}

%%
%% The abstract is a short summary of the work to be presented in the
%% article.extract data position information by fixed rules, which are unaware of the data distribution and may take long time in searching the index and tuple blocks   And existing learned indexes can support some of the common operations like bulk loading, tuple lookup, position search, tuple insert, structural modification. 

\begin{abstract}
Large language model (LLM) agents are evolving from conversational assistants into autonomous systems that execute long-horizon tasks through reasoning, tool use, code generation, and workspace manipulation. As agents increasingly operate over persistent environments and multi-step workflows, they face challenges analogous to those addressed by transactional database systems: reliable execution, consistent outcomes, safe concurrency, and durable state management. We introduce the concept of an \emph{agentic transaction} and propose an \emph{ACID-compliant agent system} framework that reinterprets the classical ACID properties for agent execution through four semantic guarantees: \textbf{Semantic Atomicity}, \textbf{Semantic Consistency}, \textbf{Semantic Isolation}, and \textbf{Semantic Durability}. Together, these properties provide a principled foundation for building reliable agent systems despite model uncertainty and dynamic execution environments. 
To instantiate this framework, we develop an ACID-compliant data agent that realizes these guarantees through transactional exploration-execution-validation cycles, transactional skill hubs, confidence divergence-based validation, semantic dependency-aware isolation, and transaction-aware semantic state management. 
Experimental results on widely used benchmarks show that our system achieves a 10.6\% improvement over state-of-the-art agents, including Claude Code. This work opens a broader research agenda on extending transactional principles and system architectures toward building trustworthy, scalable, and self-evolving AI agent systems. 
\end{abstract}

\maketitle

\iffalse                                                                                                                                                   
%%% do not modify the following VLDB block %%
%%% VLDB block start %%%
\pagestyle{\vldbpagestyle}
\begingroup\small\noindent\raggedright\textbf{PVLDB Reference Format:}\\
\vldbauthors. \vldbtitle. PVLDB, \vldbvolume(\vldbissue): \vldbpages, \vldbyear.\\
\href{https://doi.org/\vldbdoi}{doi:\vldbdoi}
\endgroup
\begingroup
\renewcommand\thefootnote{}\footnote{\noindent
This work is licensed under the Creative Commons BY-NC-ND 4.0 International License. Visit \url{https://creativecommons.org/licenses/by-nc-nd/4.0/} to view a copy of this license. For any use beyond those covered by this license, obtain permission by emailing \href{mailto:info@vldb.org}{info@vldb.org}. Copyright is held by the owner/author(s). Publication rights licensed to the VLDB Endowment. \\
\raggedright Proceedings of the VLDB Endowment, Vol. \vldbvolume, No. \vldbissue\ %
ISSN 2150-8097. \\
\href{https://doi.org/\vldbdoi}{doi:\vldbdoi} \\
}\addtocounter{footnote}{-1}\endgroup
%%% VLDB block end %%%

%%% do not modify the following VLDB block %%
%%% VLDB block start %%%
\ifdefempty{\vldbavailabilityurl}{}{
\vspace{.3cm}
\begingroup\small\noindent\raggedright\textbf{PVLDB Artifact Availability:}\\
The source code, data, and/or other artifacts have been made available at \url{\vldbavailabilityurl}.
\endgroup
}
%%% VLDB block end %%%
\fi

\vspace{-1em}

\section{Introduction}
Recent advances in large language models (LLMs) have demonstrated strong capabilities in instruction following, planning, reasoning, coding, tool use, and data processing~\cite{DBLP:journals/corr/abs-2507-01599,sun2026agenticdatabench,DBLP:journals/corr/abs-2602-02276,DBLP:journals/pvldb/HuangLZZYLZCCL25,sun2025d,zhou2024db,zhou2024d,sun2025r}.
As a result, LLM usage is shifting from single-round conversational interactions toward long-horizon production tasks, where agents operate over repository-level workspaces and autonomously coordinate iterative reasoning, code execution, and feedback-driven refinement over extended periods.
We refer to such a multi-round task-centered interactions between LLMs and execution environments as an \textit{agentic transaction} (see \autoref{fig:example}), where execution operates over semantic task states rather than structured database states.
Although agent systems differ fundamentally from databases, they face analogous challenges in ensuring {\it reliable execution, consistent outcomes, safe concurrency, and persistent state management}. Motivated by these parallels, we envision an \emph{ACID-Compliant Agent System} that reinterprets the classical ACID properties~\cite{DBLP:journals/csur/HarderR83} for agentic transactions:

\begin{figure}[!t]
\centering
\includegraphics[width=.95\linewidth]{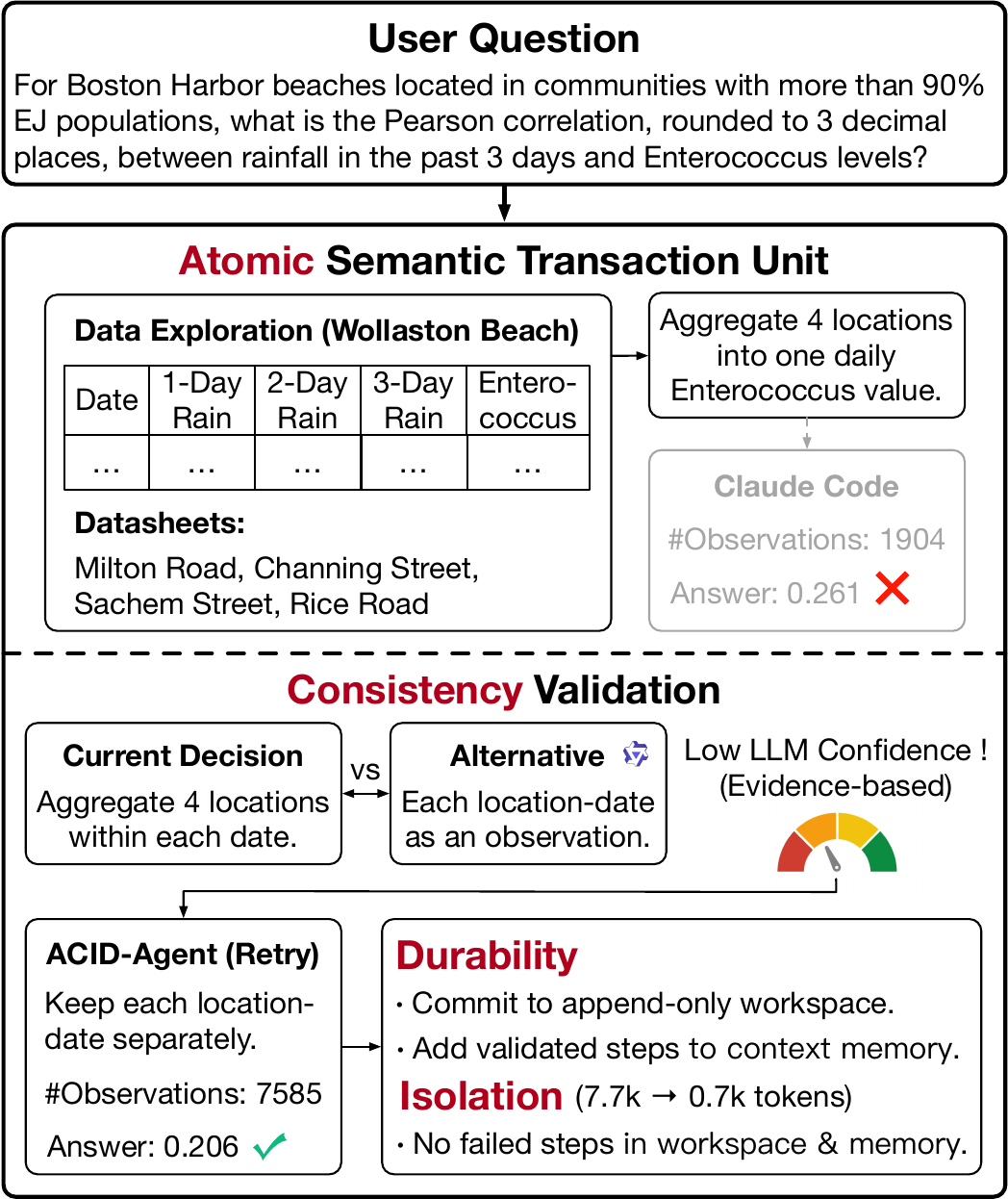}
\vspace{-1em}
\caption{An Example of ACID-Compliant Data Agent.}
\label{fig:example}
\vspace{-1.5em}
\end{figure}

\begin{table*}[!t]\vspace{-1em}
\centering
\small
\caption{ACID properties for agentic transactions. Unlike conventional
database transactions, agent transactions combine non-deterministic
reasoning with heterogeneous, potentially non-transactional effects.
The proposed semantics constrain committed effects rather than requiring
deterministic execution traces.}
\label{tab:agent-acid}
\vspace{-1em}
\renewcommand{\arraystretch}{1.18}
\begin{tabular}{
    p{0.09\textwidth}
    p{0.27\textwidth}
    p{0.31\textwidth}
    p{0.25\textwidth}}
\hline
\textbf{Property} &
\textbf{Agent-Transaction Semantics} &
\textbf{Systems Challenge} &
\textbf{Techniques} \\
\hline

\textbf{Atomicity} &
A dependency-aware set of model
invocations, tool calls, document mutations, and external actions.
Its effects become visible only if all required operations and
postconditions succeed; otherwise, all recoverable effects are
rolled back or compensated. &
Agent workflows are long-running, dynamically generated, and may
invoke external tools. Failures can
therefore leave partially updated workspaces, duplicated external
actions, or outputs unsupported by completed execution. &
Treating each exploration-execution-validation cycle as a semantic transaction unit with commit-or-retry semantics, validated effect-only commits, and test-driven transactional skill hubs.
\\
\hline

\textbf{Consistency} &
The execution
trace may be non-deterministic, but its committed outcome must satisfy
the transaction's preconditions, postconditions, and evidence
obligations. &
LLM-generated plans can be syntactically executable yet semantically
invalid because of incorrect tool selection, unsupported claims,
schema or policy violations, stale observations, and divergence
between the user's intent and the committed outcome. &
A confidence-based validation mechanism that integrates multiple reliability signals, including execution errors, decision/code confidence divergence, and LLM-based reflection feedback, with materialized skill reuse.
\\
\hline

\textbf{Isolation} &
Concurrent agent transactions must not observe or produce
semantically invalid interference. Their committed effects should be
equivalent to an execution permitted by a declared isolation level,
while allowing safe information sharing and collaboration. &
Conflicts extend beyond reads and writes to encompass prompts, memory, intermediate artifacts, tool budgets, external side effects, and derived semantic state. Moreover, conflicts often cannot be identified in advance because agents discover resources dynamically during execution. &
Isolating agent and operation contexts through dependency-aware isolation policies, isolated environments, versioned workspaces, and validation-based state control.
\\
\hline

\textbf{Durability} &
Once committed, a transaction’s effects, evidence, and recovery metadata persist across failures, enabling its state to be reconstructed and audited independently of the transient LLM context. &
Agent state is distributed across conversation context, model
outputs, tool responses, files, databases, and external services.
Model and prompt evolution complicate deterministic replay and long-term interpretation of prior executions. &
Maintaining transaction-aware memory and append-only workspaces through LLM-managed knowledge-graph evolution, provenance tracing, and version-aware recovery. 
\\
\hline
\end{tabular}
\vspace{-1.5em}
\end{table*}

\hi{Semantic Atomicity.}
Similar to how database systems encapsulate recurring application logic into reusable procedures and transaction abstractions, agent systems increasingly rely on reusable skills that package tools, workflows, and domain knowledge into coherent operational units~\cite{claudeskill}. We define \emph{semantic atomicity} as the property that an agent transaction treats a dependency-aware sequence of model invocations, tool calls, document mutations, and external actions as a single semantic unit of execution: its effects become visible only after required operations and validations succeed; otherwise, recoverable effects are rolled back or compensated.
The challenge is that agent workflows are long-running, dynamically generated, and often involve non-transactional external resources, where partial execution may leave inconsistent workspaces or unsupported results~\cite{chang2025sagallm,mohammadi2026atomix}.
For example, as shown in \autoref{fig:example}, conventional coding agents such as Claude Code may propagate intermediate decisions directly. In contrast, \oursys treats each exploration-execution-validation cycle as a semantic transaction, and only \emph{validated updates} are committed and propagated to subsequent steps.

\hi{Semantic Consistency.}
Similar to consistency guarantees in databases, agent systems must ensure that committed results remain aligned with intended semantics despite potentially non-deterministic execution~\cite{DBLP:journals/tois/HuangYMZFWCPFQL25}. We define \emph{semantic consistency} as the property that an agent transaction produces outcomes that satisfy task objectives, execution constraints, and available evidence, even when intermediate reasoning traces vary across executions.
This property is challenging because LLM-driven agents may generate syntactically executable plans that are semantically invalid due to incorrect tool selection, unsupported claims, stale observations, or divergence between user intent and execution results.
For example, as shown in \autoref{fig:example}, our \oursys employs a confidence-based validation mechanism that integrates multiple reliability signals, including execution errors, divergences in decision and code confidence, and feedback from LLM-based reflection. The mechanism grounds its assessments in evidence gathered during exploration and triggers refinement whenever confidence falls below a threshold. 

\hi{Semantic Isolation.}
Similar to concurrency control in database systems, modern agent systems increasingly execute multiple sub-agents in parallel to solve complex tasks~\cite{DBLP:journals/corr/abs-2602-02276}. We define \emph{semantic isolation} as the property that concurrent agent transactions do not observe or produce semantically invalid interference: their committed effects should be equivalent to an execution permitted by a declared isolation policy, while still allowing safe information sharing and collaboration.
For example, as shown in \autoref{fig:example}, \oursys isolates intermediate workspace states and interaction histories from failed retries, ensuring that unsuccessful attempts do not propagate to subsequent executions or agent memory.
The challenge is that agent conflicts extend beyond traditional data accesses to prompts, memories, intermediate artifacts, tool budgets, external side effects, and derived semantic states, while resource dependencies are often discovered dynamically during execution.

\hi{Semantic Durability.}
Similar to how database systems preserve transactional state and AI-native databases leverage historical transactions to improve future execution~\cite{DBLP:journals/tkde/ZhouCLS22,DBLP:journals/corr/abs-2505-22101,DBLP:journals/corr/abs-2507-01599}, agent systems must preserve semantic state beyond individual executions. We define \emph{semantic durability} as the ability to maintain committed execution states, supporting evidence, and recovery metadata beyond the lifetime of a transaction, enabling future executions to reconstruct and interpret prior results independently of transient LLM contexts.
For example, as shown in \autoref{fig:example}, \oursys maintains an append-only workspace that records committed transaction states, updating memory only with validated execution units while discarding failed attempts.
This property is challenging because agent state is distributed across conversations, model outputs, tool interactions, files, databases, and external services, while evolving models and prompts complicate reliable replay and long-term interpretation.

\begin{figure*}[!t]
  \centering
  \includegraphics[width=\linewidth]{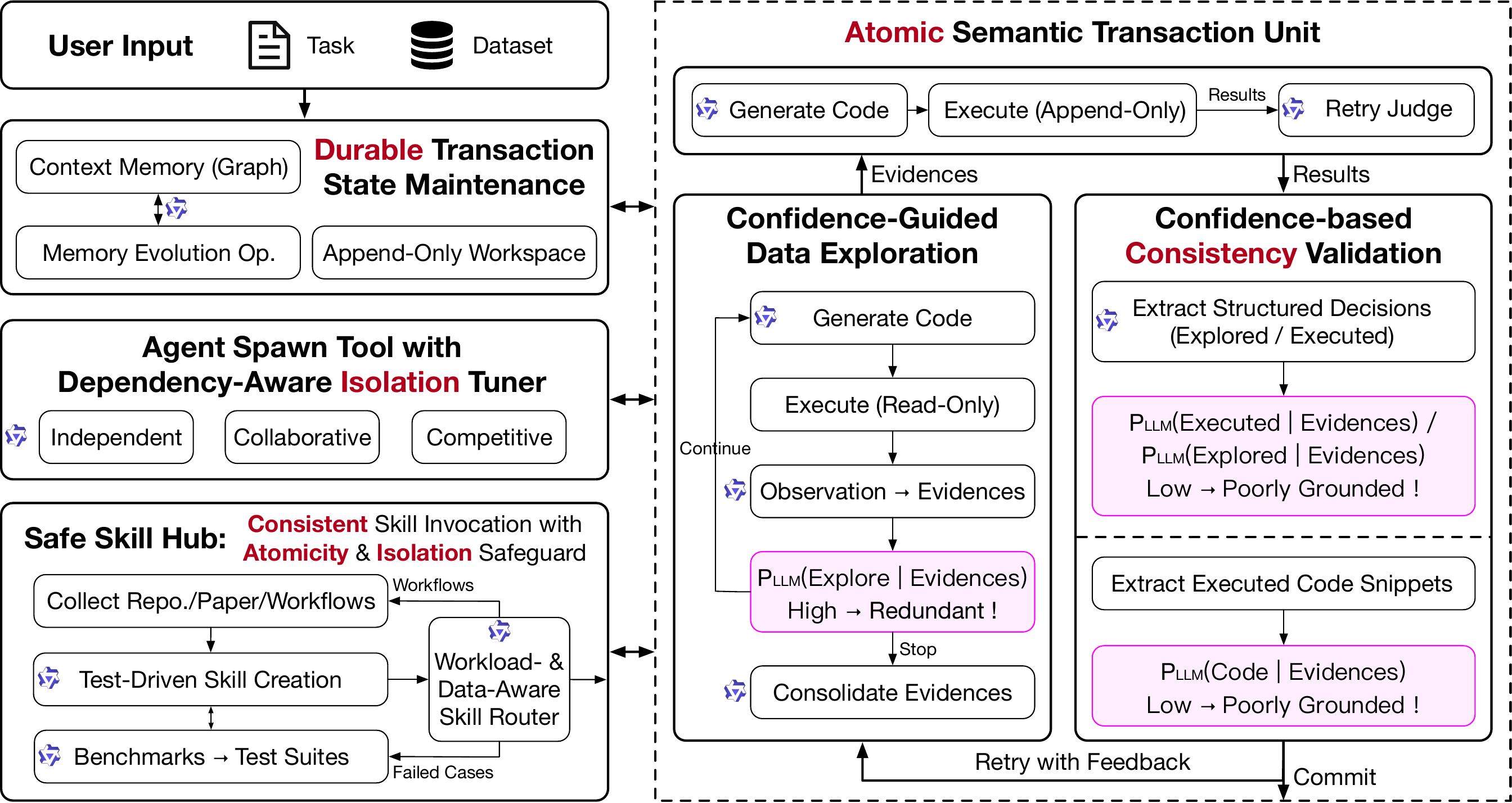}
  \vspace{-1.5em}
  \caption{Overview of ACID-Compliant Data Agent System.}
  \label{fig:data-agent}
  \vspace{-1em}
\end{figure*}

\hi{Contributions.}
In summary, we make the following contributions:
\noindent (1) We introduce a novel concept of \emph{agentic transactions} and an \emph{ACID-compliant agent system framework}, which extends the classical ACID properties to agent execution and provides a principled foundation for designing reliable agent systems.

\noindent (2) We propose an ACID-compliant data agent system:
\textbf{(A)} introduces semantic atomicity by modeling exploration-execution-validation cycles as transaction units with commit-or-retry semantics, supported by transactional skill hubs and staged execution;
\textbf{(C)} ensures semantic consistency through confidence divergence-based validation of critical decisions and generated code, integrating execution signals and LLM feedback to detect unsupported behaviors and trigger evidence-guided retries;
\textbf{(I)} enables semantic isolation by regulating dependencies among agents, contexts, and operations through adaptive coordination strategies, isolated execution environments, and versioned workspaces;
\textbf{(D)} achieves semantic durability through transaction-aware memory, append-only workspace management, and persistent execution traces for recovery and long-horizon reasoning (see Section \ref{sec:overview}).

\noindent (3) Our preliminary experimental results demonstrate the potential advantages of transactionally designed agent systems (see Sections~\ref{sec:exp}).
We also provide open research problems for extending agentic transactions to the full lifecycle of agent systems (see Section~\ref{sec:roadmap}).
The source code is available at \url{https://github.com/TsinghuaDatabaseGroup/ACID-Agent}.

\section{Agentic Transaction} \label{sec:overview}

\subsection{Preliminaries}
\label{subsec:pre}
LLM-based agents can solve tasks through long-horizon cycles of LLM reasoning, tool/skill invocation, and execution feedback.
We refer to such multi-round task-centric interactions between LLMs and environments as \textit{agentic transactions}.

\begin{definition}[Agentic Transaction]
An agentic transaction, $\tau$, is a bounded unit of agent execution comprising a finite sequence of LLM-driven interactions between an agent and its execution environment, undertaken to accomplish a task. Formally, given a tool set $T$ and a skill set $S$, an agentic transaction $\tau = \langle r_1, \dots, r_n \rangle$ comprises $n$ steps. Each step $r_i = (c_i, a_i, f_i)$ consists of an LLM context $c_i$ (including the tools in $T$ and skills in $S$ available at that step), an agent action $a_i$ that invokes a tool or skill, and the resulting feedback $f_i$ from the environment. The transaction $\tau$ commits only if its execution satisfies the required task conditions and preserves all semantic invariants. Otherwise, its intermediate effects are rolled back or compensated for to ensure that no invalid side effects remain.
\end{definition}

For example, \autoref{fig:example} presents a data-agent transaction. It begins with LLM-driven exploration of datasets and schemas, followed by iterative analysis steps that invoke tools, update the workspace, and refine decisions based on feedback. Each exploration-execution-validation cycle forms a semantic transaction unit whose effects are propagated only after validation. Failed units are discarded or recovered without affecting the committed state.

\autoref{tab:agent-acid} summarizes representative challenges and techniques for achieving the ACID properties of agentic transactions.

\subsection{An ACID-Compliant Data Agent System}
\label{sec:data-agent}

Data agents aim to automate data science workflows, derive insights from heterogeneous data, and manage data systems.
Based on agentic transaction principles, we propose an \emph{ACID-compliant data agent system}, as shown in \autoref{fig:data-agent}, providing reliability guarantees for exploration, execution, and transactional state evolution.

\begin{figure*}[!t]
  \centering
  \includegraphics[width=\linewidth]{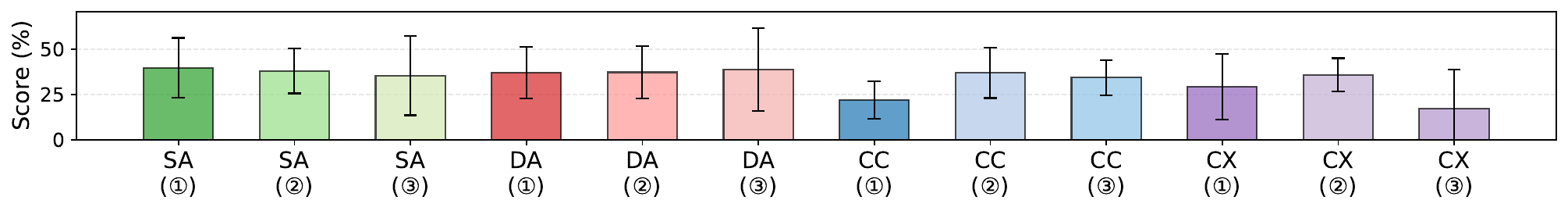}
  \vspace{-2.5em}
  \caption{Consistency of Agent Performance Across Three Runs on 10 \textit{AgenticDataBench}~\cite{sun2026agenticdatabench} Tasks. Bars show mean performance; error bars show the square root of the average per-task variance across runs. SA=Smolagents, DA=DA-Agent, CC=Claude Code, CX=CodeX. \ding{192}=Qwen3.5-397B-A17B, \ding{193}=Kimi-K2.5, \ding{194}=Claude Sonnet 4.6.}
  \label{fig:pass3}
  \vspace{-1.5em}
\end{figure*}

\subsubsection{Semantic Atomicity}
\label{subsec:atomicity}

We propose two mechanisms for semantic atomicity: an offline skill hub that embeds transactional safeguards into reusable skills, and an online staged-execution framework that enforces commit-or-retry semantics via validation gates.

\hi{Offline Skill Hub Creation.}
Under semantic atomicity, tools and skills become first-class transactional objects with lifecycle interfaces, and only validated effects are committed.  We envision targeted skill hubs that automatically enforce transactional semantics, including rollback and commit, without requiring agents to perform manual transaction management. First, for workspace-modifying skills, this requires preventing invalid partial updates and side effects through idempotency keys, write-ahead action logs, checkpointing, and automatic compensation.
Second, for system optimization skills, it further requires avoiding regressions caused by conflicting skill interactions.
Realizing such skill hubs requires distilling existing data management expertise into deployable agent skills with adaptive routing and built-in transactional guarantees.
We address this by packaging existing repositories as agent skills with standardized CLIs and validating their behavior through LLM-generated test suites derived from established benchmarks. 
We further develop a workload- and data-aware skill router that periodically analyzes historical system logs, uses LLM to summarize evolving workload characteristics, and dynamically adjusts feature importance for both skill retrieval and LLM-based skill selection.

\hi{Online Semantic Transaction Execution.}
When existing skills with built-in safeguards are unavailable, we enforce semantic atomicity through an online semantic transaction framework.
Specifically, we model the agent trajectory as a sequence of exploration-execution-validation cycles, with each cycle treated as a semantic transaction unit governed by commit-or-retry semantics.
First, \emph{confidence-guided data exploration} enables agents to iteratively collect and consolidate evidence while avoiding redundant exploration. 
Second, \emph{confidence-based consistency validation} verifies the reliability of agent decisions and generated code by integrating execution errors, confidence divergence, and LLM-based reflection signals. Violations beyond predefined thresholds trigger retry.
The failed execution steps are discarded by isolating their intermediate contexts from memory and excluding their workspace updates, ensuring atomic state evolution through an append-only workspace.

We first describe confidence-guided data exploration and defer the second component to Section \ref{subsec:consistency}.
Specifically, given a task, the exploration sub-agent iteratively generates read-only exploration code using the task description and recent exploration summaries as context.
After execution, the observations are summarized into exploration memory.
To avoid redundant exploration, we use LLM confidence-based validation to compare the current exploration observation with and without previous observations as context (detailed in Section \ref{subsec:consistency}).
A large confidence divergence indicates that the current exploration is heavily dependent on prior observations and provides limited new information.
We terminate exploration when redundant exploration exceeds a predefined threshold.
Exploration summaries are further consolidated to resolve potential conflicts and maintain a consistent evidence base.

\subsubsection{Semantic Consistency}
\label{subsec:consistency}

Data agents exhibit execution inconsistency under workflow and model uncertainties, with repeated runs showing substantial variance and occasional intent violations (see \autoref{fig:pass3}). 
Thus, we develop complementary offline and online mechanisms. Offline, we reuse validated workflows and enhance model-level behavioral stability. Online, we detect and correct execution deviations through trajectory-level consistency validation.

\hi{Offline Consistency Enhancement.}
$(i)$ \emph{Workflow-Level Materialization.}
Similar to materialized views in databases, successful execution workflows can be materialized into reusable agent skills with semantic validation logic. These skills are retrieved based on task semantics and execution context, with details discussed in Section \ref{subsec:atomicity}.
$(ii)$ \emph{Model-Level Stability.}
Consistency also depends on model reliability. Existing benchmarks focus on single-run correctness or best-of-N performance while overlooking execution stability. We envision consistency-oriented benchmarks to measure stability and guide targeted fine-tuning.

\hi{Online Consistency Validation.}
To maintain execution consistency with task requirements and supporting evidence, we propose a confidence-based validation mechanism that integrates multiple reliability signals, including execution errors, decision/code confidence divergence, and LLM-based reflection feedback. A retry is triggered when any signal exceeds its predefined threshold, with the validation feedback incorporated into the agent context to guide subsequent exploration-execution attempts. The process terminates upon successful validation or reaching the retry limit.

To operationalize confidence-based validation, we quantify LLM confidence as the exponential of the average token-level log probability over the target output.
Confidence divergence between two contexts is then measured by comparing their corresponding confidence scores.
We instantiate this measure for different agent outputs as follows.
$(i)$ \textit{Decision Confidence Divergence.} For critical decisions (e.g., filter predicates), we use LLMs to extract explored decisions from exploration summary and executed decisions from code.
Given the task and recent exploration summaries as context, we measure their confidence divergence; a low divergence indicates that the executed decision does not gain stronger evidence support than alternative explored decisions.
$(ii)$ \textit{Code Confidence Divergence.} For code generation, we identify decision-relevant code spans through static code analysis (e.g., control flows), and evaluate their confidence with and without exploration evidence.
A low confidence divergence indicates that the generated code is insufficiently grounded in supporting evidence and warrants inspection.

\subsubsection{Semantic Isolation}
Unlike traditional transactions that mainly isolate conflicting data accesses, agentic transactions require isolation over semantic dependencies among agents, contexts, workspaces, and operations. We consider two levels of isolation: agent–agent isolation and operation–operation isolation.

\hi{Agent-Level Isolation.}
Different dependency structures among sub-agents introduce distinct coordination requirements, motivating adaptive isolation strategies supported by agent spawning tools. We formulate isolation selection as a semantic parameter tuning problem, where isolation policies (e.g., access constraints, branching strategies, communication intervals, and termination conditions) are determined based on sub-task semantics and dependencies. This enables learning-based optimization of isolation policies, where a dedicated LLM can be fine-tuned to predict suitable configurations for different tasks, analogous to LLM-based database knob tuning~\cite{DBLP:journals/pvldb/HuangLZZYLZCCL25}.
Specifically, $(i)$ \emph{Independent sub-agents} address semantically disjoint sub-tasks (e.g., summarizing large collections of independent documents), where each sub-agent operates on an exclusive subset of resources with isolated permissions and fully parallel execution.
$(ii)$ \emph{Collaborative sub-agents} jointly construct a shared artifact (e.g., soft engineering involving coordinated code evolution and integration over a shared codebase), where agents maintain independent workspace branches, periodically synchronize intermediate results through structured context exchange, and merge changes using a Git-like workflow.
$(iii)$ \emph{Competitive sub-agents} explore alternative hypotheses or solution strategies (e.g., conducting in-depth research by analyzing related literature, validating claims, and synthesizing evidence), where each agent executes in an isolated virtual environment (e.g., Docker) and the final result is selected from the most promising trajectory.
Efficiency can be further improved through copy-on-write initialization and early termination of under-performing branches.

\hi{Operation-Level Isolation.}
At the operation level, the skill hub enforces execution isolation through effect annotations and inference, versioned workspaces, snapshot-based execution, and optimistic validation, preventing failed or conflicting operations from contaminating shared states.

\subsubsection{Semantic Durability}
We provide semantic durability through $(i)$ transaction-aware semantic state management during execution, and 
$(ii)$ persistent execution tracing and recovery.

\hi{Transaction-Aware Semantic State Management.} 
Unlike traditional databases that maintain explicit structured states, AI systems operate over evolving semantic states generated from interactions, workspace updates, tool executions, and intermediate artifacts. Preserving these states is critical for downstream reasoning, but growing interaction histories quickly exceed finite context windows.
Existing approaches typically rely on step-wise LLM summarization to compress interaction histories.
However, this process often fails to preserve transaction-level semantic structure, either over-compressing critical information or retaining irrelevant step-level details that are not useful for downstream reasoning.
To address this challenge, we propose a transaction-aware evolving memory that maintains semantic states throughout the transaction lifecycle. The memory evolves as a knowledge graph, with insertion, merging, splitting, and deletion operations performed by a specialized LLM.
Training supervision can be automatically derived from agent trajectories, where information referenced by future execution steps serves as a signal of long-horizon relevance.

\hi{Execution Tracing and Recovery.}
Beyond maintaining semantic states during execution, we preserve durable execution histories to support auditing and recovery. Our append-only workspace records provenance information, LLM interactions, tool invocations, and versioned artifacts throughout the transaction lifecycle. These traces enable faithful reconstruction of execution environments, diagnosis of failures, and version-aware failure recovery~\cite{dong2026deltabox}.

\section{Experiments}\label{sec:exp}

\subsection{Experimental Setup}
All experiments are conducted on a Linux server with 256GB RAM, Intel(R) Xeon(R) Silver 4110 CPU @ 2.10GHz CPU, and NVIDIA GeForce RTX 2080 Ti.
Although agentic transactions represent a broader concept, we first validate our ACID-compliant data agent system as an initial proof of effectiveness.

\hi{Dataset.} We use KramaBench~\cite{lai2025kramabench}, a representative benchmark for data agents.
It contains 104 natural language tasks over 1,700 real-world data files collected from 24 data sources across 6 domains.
Each task specifies a data science objective, requires reasoning over heterogeneous datasets, and involves multi-step workflows.

\hi{Evaluated Methods.}
We evaluate state-of-the-art LLMs, including Qwen3.5-397B-A17B~\cite{qwen3.5} and GLM-5.2~\cite{glm5.2}, provided by the Bailian platform~\cite{bailian}. We use default temperatures.
We evaluate three representative data-agent harnesses: $(i)$ \claudecode~\cite{claudecode}, a general-purpose ReAct-style harness with long-horizon planning, environment interaction, and context management; 
$(ii)$ \oursys, using a local Qwen3-0.6B~\cite{qwen306b} for confidence estimation since API-based LLMs lack token probabilities.
We set the maximum number of semantic units to 20, retain up to 15 historical units, and allow 2 retries per unit. Exploration uses an adaptive budget of 1–4 rounds, decreasing by one round every two units, and terminates early when confidence divergence exceeds 0.45. Retries are triggered when decision confidence divergence is below 0.25 or the maximum code-span confidence divergence is below 0.50;
$(iii)$ \daagent~\cite{huang2024code}, a data science agent with Bash, Python, and SQL tools for reactive execution with feedback, serving as an ablation variant of \oursys that removes ACID designs.

\begin{table*}[!t]
\caption{Scores (\%) and Trajectory-level Metrics on KramaBench. \ding{192}=Qwen3.5-397B-A17B, \ding{193}=GLM-5.2.}
\vspace{-1em}
\label{tab:overall}
\resizebox{\linewidth}{!}{
\begin{tabular}{clcccccccccc}
\hline
\multirow{2}{*}{\textbf{Harness}} & \multirow{2}{*}{\textbf{LLM}} & \multirow{2}{*}{\textbf{Score}} & \multicolumn{6}{c}{\textbf{Domain Scores}} & \multirow{2}{*}{\textbf{\#Code Steps}} & \multirow{2}{*}{\textbf{\#Tokens (K)}} & \multirow{2}{*}{\textbf{Cost (\$)}} \\ \cline{4-9}
& & & Archaeology & Astronomy & Biomedical & Environment & Legal & Wildfire & & & \\ \hline
\multirow{2}{*}{\claudecode} & \ding{192} Qwen & 64.0 & 41.7 & 54.2 & 44.4 & 70.2 & 73.3 & 71.7 & 9.4 & 405 & \textbf{0.08}  \\ \cline{2-12}
& \ding{193} GLM & 74.2 & \textbf{50.0} & 54.2 & 55.6 & 90.0 & 80.0 & \textbf{84.2} & \textbf{8.8} & \textbf{289} & 0.12 \\ \hline
\multirow{2}{*}{\oursys} & \ding{192} Qwen & 74.6 & 41.7 & \textbf{58.3} & 55.6 & 90.0 & \textbf{83.3} & 83.7 & 22.8 & 348 & 0.10 \\ \cline{2-12}
& \ding{193} GLM & \textbf{77.4} & \textbf{50.0} & \textbf{58.3} & \textbf{77.8} & \textbf{95.0} & 80.0 & 83.1 & 22.5 & 367 & 0.61 \\ \hline
\end{tabular}
}
\vspace{-1em}
\end{table*}

\hi{Evaluation Metrics.} We evaluate agents from three perspectives: task quality, execution efficiency, and result consistency.
Task quality is measured by the benchmark score, while efficiency is evaluated by trajectory-level statistics, including coding steps, token consumption, and execution cost.
Consistency is measured by the square root of the average per-task variance across multiple runs.

\subsection{Main Results}
We evaluate \oursys on KramaBench from two perspectives: overall performance and execution consistency.
We measure task scores and quantify consistency using the average per-task variance across three independent runs.

\hi{Overall Evaluation.} As shown in \autoref{tab:overall}, \oursys consistently achieves higher overall scores than \claudecode across different LLM backbones, with improvements observed in most domains.
Powered by Qwen3.5-197B-A17B, \oursys outperforms \claudecode by 10.6\% in overall score.
Moreover, \oursys with Qwen3.5-397B-A17B even surpasses \claudecode with the larger GLM-5.2 backbone, demonstrating the effectiveness of our harness design beyond model scaling.
This improvement comes at the cost of additional code steps and token consumption, mainly due to exploration and retry mechanisms.
The results demonstrate that leveraging a lightweight local model (Qwen3-0.6B) for consistency validation effectively complements much stronger backbone LLMs.

\hi{Consistency Evaluation.}
As shown in \autoref{tab:consistency}, \oursys achieves lower task-level score variation than \claudecode, indicating that confidence-guided exploration and validation can mitigate the non-deterministic deviations from transactional semantics.

\begin{table}[!t]
\caption{Consistency of Agent Performance Across Three Runs on the Environment Domain of KramaBench ($\mathrm{mean}\pm\sqrt{\mathrm{avg}(\mathrm{Var})}$). All agents are based on Qwen3.5-397B-A17B.}
\vspace{-1em}
\label{tab:consistency}
\resizebox{\linewidth}{!}{
\begin{tabular}{ccccc}
\hline
\textbf{Data Agent} & \textbf{Score} & \textbf{\#Code Steps} & \textbf{\#Tokens (K)} & \textbf{Cost (\$)} \\ \hline
\claudecode & 63.9 $\pm$ 30.9 & \textbf{8.5 $\pm$ 2.9} & \textbf{372 $\pm$ 161} & \textbf{0.07 $\pm$ 0.03}\\ \hline
\oursys & \textbf{88.9 $\pm$ 18.6} & 25.1 $\pm$ 9.8 & 421 $\pm$ 200 & 0.13 $\pm$ 0.06\\ \hline
\end{tabular}
}
\vspace{-1em}
\end{table}

\subsection{Ablation Study}
The ablation results are shown in \autoref{tab:ablation}.
$(i)$ \emph{Semantic Transaction Unit}. 
\oursys outperforms \daagent, demonstrating the effectiveness of exploration-execution-validation cycles compared with conventional ReAct-style execution.
$(ii)$ \emph{Failed Step Isolation}. 
We remove the isolation mechanism for failed steps, allowing intermediate failures to directly update the workspace and context memory. This variant reduces the score by 11.7\%, demonstrating that propagating failed states can contaminate subsequent execution.
$(iii)$ \emph{Effect of More Tokens}. 
\oursys outperforms majority-voting \claudecode across three runs with fewer tokens. This indicates that the improvement does not come from increased inference budgets, but from the ACID-inspired harness design.

\begin{table}[!t]
\caption{Ablation Study on the Environment Domain of KramaBench. All agents are based on Qwen3.5-397B-A17B.}
\vspace{-1em}
\label{tab:ablation}
\resizebox{\linewidth}{!}{
\begin{tabular}{ccccc}
\hline
\textbf{Data Agent} & \textbf{Score} & \textbf{\#Code Steps} & \textbf{\#Tokens (K)} & \textbf{Cost (\$)} \\ \hline
\daagent & 65.2 & 8.5 & 62 & 0.01 \\ \hline
\begin{tabular}[c]{@{}c@{}}\claudecode\\(3-Majority)\end{tabular} & 75.2 & 25.6 & 1121 & 0.21\\ \hline
\begin{tabular}[c]{@{}c@{}}\oursys\\(No-Isolation)\end{tabular} & 78.3 & 20.1 & 333 & 0.10 \\ \hline
\oursys & 90.0 & 25.5 & 444 & 0.13\\ \hline
\end{tabular}
}
\vspace{-1.25em}
\end{table}
\section{Open Problems}
\label{sec:roadmap}

Beyond ACID-compliant data agents, building ACID-compliant general-purpose agentic systems raises several open research questions. 
For \textbf{atomicity}, how can we build scalable skill ecosystems with executable semantics, quality assurance, and safety guarantees that support reliable skill composition and evolution? 
For \textbf{consistency}, how can we ensure reliable reasoning and stable execution across runs through new architectures, benchmarks, model alignment techniques, execution harnesses, machine-checkable contracts, and typed tool interfaces? 
For \textbf{isolation}, how can multi-agent systems safely coordinate access to shared contexts, tools, artifacts, and semantic states? Addressing this challenge requires transactional abstractions and integrated database--LLM serving mechanisms for managing context, ownership, and conflicts~\cite{kang2026thunderagent,seekdb}. 
For \textbf{durability}, how can we turn agent memory into systematic infrastructure that supports transactional state management, failure recovery, and persistent evolution for lifelong agents?

% \clearpage

% \balance
\bibliographystyle{ACM-Reference-Format}
\bibliography{refs}
% \balance

\end{document}